\documentclass[letter]{aa} 

\usepackage{graphicx}	
\usepackage{amsmath}	
\usepackage{amsmath, amsfonts, amssymb, graphics,graphicx, wrapfig, nicefrac}
\usepackage{float}
\usepackage{epsfig}
\usepackage{epstopdf}
\usepackage{multirow}
\usepackage{multicol}
\usepackage{graphicx}
\usepackage{rotating}
\usepackage{mathrsfs}
\usepackage{stfloats}
\usepackage{capt-of}
\graphicspath{{./}}
\usepackage{txfonts}
\usepackage[colorlinks=true,linkcolor=blue, citecolor=blue,urlcolor=magenta]{hyperref}
\begin{document}

   \title{Anomalously high deuterium fractionation in a galactic translucent cloud: a challenge to chemical models}


   \author{Gan Luo
          \inst{1,2}
          \and
          Zhi-Yu Zhang\inst{2}
          \and
          Thomas G. Bisbas\inst{3}
          \and
          Di Li\inst{4,5}
          \and
          Serena Viti\inst{6,7,8}
          \and 
          Roberto Neri\inst{1}
          \and
          Junzhi Wang\inst{9}
          \and
          Siyi Feng\inst{10}
          \and
          Ningyu Tang\inst{11}
          \and
          Daniel R. Rybarczyk\inst{12}
          \and
          Lingrui Lin\inst{2}
          }

   \institute{Institut de Radioastronomie Millimétrique, 300 rue de la Piscine, 38400, Saint-Martin d’Hères, France
        \email{luo@iram.fr}
        \and
            School of Astronomy and Space Science, Nanjing University, Nanjing 210093, People’s Republic of China
        \and
            Research Center for Computational Earth and Space Science, Zhejiang Lab, Hangzhou, 311100, PR China
        \and
            New Cornerstone Science Laboratory, Department of Astronomy, Tsinghua University, Beijing 100084, China
        \and
            State Key Laboratory of Radio Astronomy and Technology, National Astronomical Observatories, Chinese Academy of Sciences, Beijing 100101, China
        \and
            Leiden Observatory, Leiden University, PO Box 9513, NL-2300 RA Leiden, The Netherlands
        \and
            Transdisciplinary Research Area (TRA Matter) and Argelander Institut für Astronomie, Universität Bonn, Auf dem Hügel 71, D-53121 Bonn, Germany
        \and
            Department of Physics and Astronomy, UCL, Gower Place, London WC1E 6BT, UK
        \and
            School of Physical Science and Technology, Guangxi University, Nanning 530004, People's Republic of China
        \and
            Department of Astronomy, Xiamen University, Zengcuo'an West Road, Xiamen, 361005, People’s Republic of China
        \and
            Department of Physics, Anhui Normal University, Wuhu, Anhui 241002, China
        \and
            University of Wisconsin–Madison, Department of Astronomy, 475 N Charter St, Madison, WI 53703, USA
             }

   \date{Received xx; accepted xx}

\abstract{Deuterated (D-) species have long been proposed to diagnose the physical conditions and chemical evolution of cold dense molecular clouds. While deuterium fractionation has been extensively measured in dense cores, observations in diffuse and translucent clouds remain rare. We report here the detection of DCN and DNC toward a translucent cloud ($A_{\rm V} =1.2\pm0.2$~mag, $n_{\rm H_2}$ = $3.9\pm0.2\times10^2$~cm$^{-3}$) through sensitive absorption observations with the IRAM NOrthern Extended Millimeter Array (NOEMA). This detection reaches the lowest column-density and volume-density regime in which deuteration has been observed so far. Interestingly, the observed DCN/HCN and DNC/HNC abundance ratios ($3.3\pm0.6\times10^{-3}$ and $3.6\pm1.2\times10^{-3}$, respectively), which are more than two orders of magnitude higher than the element abundance [D]/[H] (1.5$\times$10$^{-5}$), suggest an unexpected enhancement of deuterium fractionation in the translucent cloud. These results represent a significant departure from established chemical models considering deuterium fractionation, which predict negligible formation of D-molecules in such environments. Although it remains unclear how D-molecules built up their abundances in translucent gas, a dispersed dense core scenario could potentially explain the observed high deuterium fraction. This interpretation is consistent with the idea proposed by \citet{Price2003} more than two decades ago: a translucent cloud may be a transient, dynamically evolving structure formed through the dissipation of a dense molecular cloud.}




   \maketitle
   \nolinenumbers
%

\section{Introduction}

Deuterated (D-) species\footnote{Throughout the paper, we refer to as deuterated molecules excluding HD.} are crucial for understanding the evolution and thermal history of molecular clouds and protoplanetary disks \citep{Aikawa2012,Ceccarelli2014,Jensen2019}. Deuterium is produced in the Big Bang and is only destroyed through nucleosynthesis in stellar interiors \citep{Epstein1976,Boesgaard1985}. While the measured element abundance of deuterium to hydrogen (D{\sc i}/H{\sc i}) is 1.5$\times$10$^{-5}$ \citep{Linsky2003,Oliveira2003,Linsky2006,Friedman2023}, the deuterium fraction ($f_{\rm D}$, the ratio between the D-species and their hydrogen counterparts) measured in molecules is usually a few orders of magnitude higher in cold ($T\lesssim20$ K), dense cores \citep{Turner2001,Caselli2002b,Bacmann2003,Gerner2015,Feng2019,Jorgensen2020,Giers2022}. 

The incorporation of deuterium in various molecules, namely deuterium fractionation, is due to the different zero-point vibration energies between H- and D-isotopologues \citep{Wilson1994}. 
This process is particularly efficient in cold environments, where the deuteration of H$_3^+$ proceeds through the exothermic reaction: ${\rm H_3^+ + HD  \rightarrow H_2D^+ + H_2 +230K}$ \citep{Watson1976, Millar1989, Caselli2002b}. Subsequent deuteration of molecules (e.g., DCO$^+$, N$_2$D$^+$) forms via proton transfer reactions involve H$_2$D$^+$ with neutral species (e.g., N$_2$, CO). In warmer environments (e.g., $T>20$~K), CH$_2$D$^+$ is believed to be an important precursor to form deuterated species such as DCN, HDCO, etc \citep{Millar1989,Turner2001,Roueff2007}. 

A recent survey of DCN and DCO$^+$ toward {\it Planck} Galactic Cold Clumps (PGCCs) suggests that the deuteration process may begin earlier, potentially already during relatively diffuse or transitory stages of cloud evolution rather than exclusively during the dense-core phase \citep{Mo2025}. 
To date, most of the D-molecules are measured in high-density environments \citep[$n_{\rm H}$ $>$10$^4$\,cm$^{-3}$,][]{Caselli2002b,Crapsi2005,Minh2018,Feng2019,Feng2020,Navarro-Almaida2021}. As the current physical prescriptions in the literature stand, the deuterium fractionation has, however, not been reported yet in Galactic diffuse and translucent clouds \citep[$n_{\rm H} \leq {\rm a \ few} 10^3$\,cm$^{-3}$, $A_{\rm V}<5$~mag,][]{Snow2006}. 

The major difficulties in measuring deuterium fraction in translucent clouds are the low molecular abundances and sub-thermal excitation of even the lowest rotational transitions \citep[$T_{\rm ex} \approx 2.73$ K,][]{Godard2010,Luo2020}. Thus, it is almost impossible to detect them in emissions in translucent clouds. Measuring absorption lines against strong continuum sources with interferometers has proven to overcome difficulties in determining the physical and chemical properties of diffuse and translucent clouds \citep{Liszt1996,Muller2011, Muller2013,Muller2020,Luo2023,Luo2024b,Rybarczyk2022b,Rybarczyk2022a,Rybarczyk2023,Liszt2023a,Liszt2023b}.

In this Letter, we report the first detection of DCN and DNC in absorption with NOEMA toward a Galactic translucent cloud in front of J0418+3801 (B0415+379), a sightline that has been extensively observed over the past 30 years. These measurements provide a sensitive probe of deuterium chemistry in low-density molecular gas. We discussed the inferred deuterium fractionation in the context of current astrochemical models and the possible implications for the origin of translucent clouds.

\section{Observations}\label{sec:obs}

We carried out NOEMA observations of $J$=1--0 transitions of HCO$^+$, HCN, HNC, and their $^{13}$C- and D-isotopologues toward J0418+3801 in December 2020, January 2021, and March and April 2026 (Project IDs: W20BB and E25AE, PI: Gan Luo). The total equivalent (12 antennas) observing time is 24.5 hours (15~hr on source) and the spectral resolution is 62.5~kHz (corresponding to a velocity resolution of 0.26 km\,s$^{-1}$ at 72~GHz). We used 3C~84 as the bandpass and phase calibrator, and LKHA101 and MWC349 as the flux calibrators. The calibration of the raw data was performed using the {\sc clic} package in {\sc gildas}\footnote{https://www.iram.fr/IRAMFR/GILDAS} \citep{Pety2005}, and the calibrated visibilities were combined for all observing tracks. The source is self-calibrated to reduce the phase calibration error caused by atmospheric phase fluctuations between the calibrator and the target. The bandpass calibrator was observed for a total of $\sim$3.7~hr to ensure that the bandpass errors do not introduce artifacts that would bias the calibration of the source's spectral profile. 
The resultant noise level of the normalized spectra (see Sect.~\ref{sec:lines}) reached $\sim2.1\times10^{-3}$ at 72~GHz and $\sim1\times10^{-3}$ at 89~GHz per 62.5~kHz channel.

\section{Results and discussion}\label{sec:results}
\subsection{Absorption profiles and Gaussian fitting}\label{sec:lines}

The continuum background is unresolved under current NOEMA observations ($\sim1.5''$, corresponding to $\sim400$~au at a distance of 300~pc); thus, we used a point function to fit the visibilities. Given the bright continuum background ($S_{\rm 72~GHz}\sim1.4$~Jy), the normalized spectra can be written as:
\begin{equation}
e^{-\tau_\nu}=T_{\rm B}/J(T_{\rm cont}),\label{eq:eq1}
\end{equation}
where $\tau_\nu$ is the optical depth at given frequency $\nu$, $T_{\rm B}$ is the brightness temperature, and $J\mathrm{(T)}$ = $(h\nu/k)/(e^{h\nu/kT}-1)$ is the Rayleigh Jeans equivalent temperature.

\begin{figure*}
\centering
\includegraphics[width=1.0\textwidth]{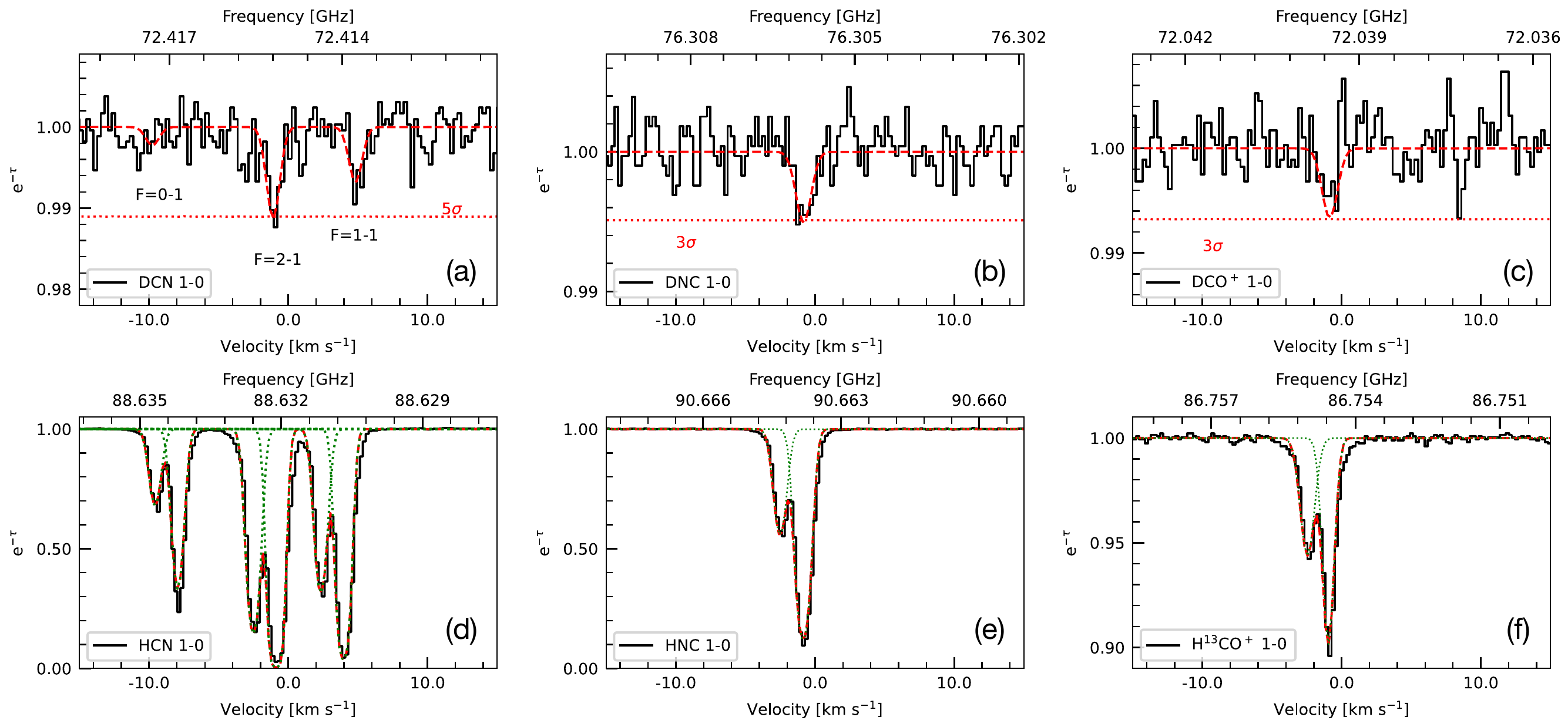}
\vspace{-2.0em}
\caption{Normalized spectra of DCN (a), DNC (b), DCO$^+$ (c), HCN (d), HNC (e), and HCO$^+$ (f) {\it J} = 1-0 transitions. The red dashed curve in each panel represents the fitting to the spectra, except for DCO$^+$, where we outline the artificial 3$\sigma$ signal (channel-based) using the same line width as H$^{13}$CO$^+$. The red dotted curves denote 5$\sigma$ noise level in (a) and 3$\sigma$ noise level in (b) and (c).
}
\label{fig:fig1}
\end{figure*}

The normalized absorption profiles and Gaussian decomposition of $J$=1--0 transitions of HCN, HNC, H$^{13}$CO$^+$, and their D-isotopologues are shown in Fig.~\ref{fig:fig1}(a)--(f). For the hyperfine transitions, we fixed the optical depth ratios to the intrinsic intensity ratios (1:3:5) in the spectra fitting. HCN, HNC, and H$^{13}$CO$^+$ show detections at both $-2.5$ and $-0.9$~km~s$^{-1}$ components, while DCN and DNC have only detection in the $-0.9$~km~s$^{-1}$ component. All the detections have consistent line central velocity and line width ($\Delta V\sim1$~km~s$^{-1}$). The optical depths and column densities are summarized in Table \ref{tab:tab1}.  
DCO$^+$ was not detected, we only give a 3$\sigma$ upper limit (channel-based). Since the gas kinetic temperature is $T_{\rm k}=55\pm2$~K (see Appendix \ref{sec:density}), the detections of DCN and DNC but not for DCO$^+$ are consistent with the fact that DCO$^+$ is formed in cold environments through the H$_2$D$^+$ channel, while the other two species can form in warm environments \citep{Roueff2007,Colzi2022}. 

\subsection{The anomalously high deuterium fraction}\label{sec:fD}

The derived abundance ratios of DCN/HCN and DNC/HNC are (3.3$\pm$0.6)$\times$10$^{-3}$ and (3.6$\pm$1.2)$\times$10$^{-3}$, respectively\footnote{The value could be slightly overestimated due to the underestimation of column density from optically thick absorptions \citep{Luo2024b}.}. The 3$\sigma$ upper limit of DCO$^+$/HCO$^+$\footnote{The column density of HCO$^+$ is estimated from H$^{13}$CO$^+$, assuming $^{12}$C/$^{13}$C = 74 \citep{Lucas1998,Luo2024b}} is 3.1$\times$10$^{-3}$. The measured deuterium fraction is more than two orders of magnitude higher than the elemental D/H ratio, suggesting that deuterium fractionation may have already played a significant role in the low-density gas. 

\begin{figure*}
\centering
\includegraphics[width=1.0\textwidth]{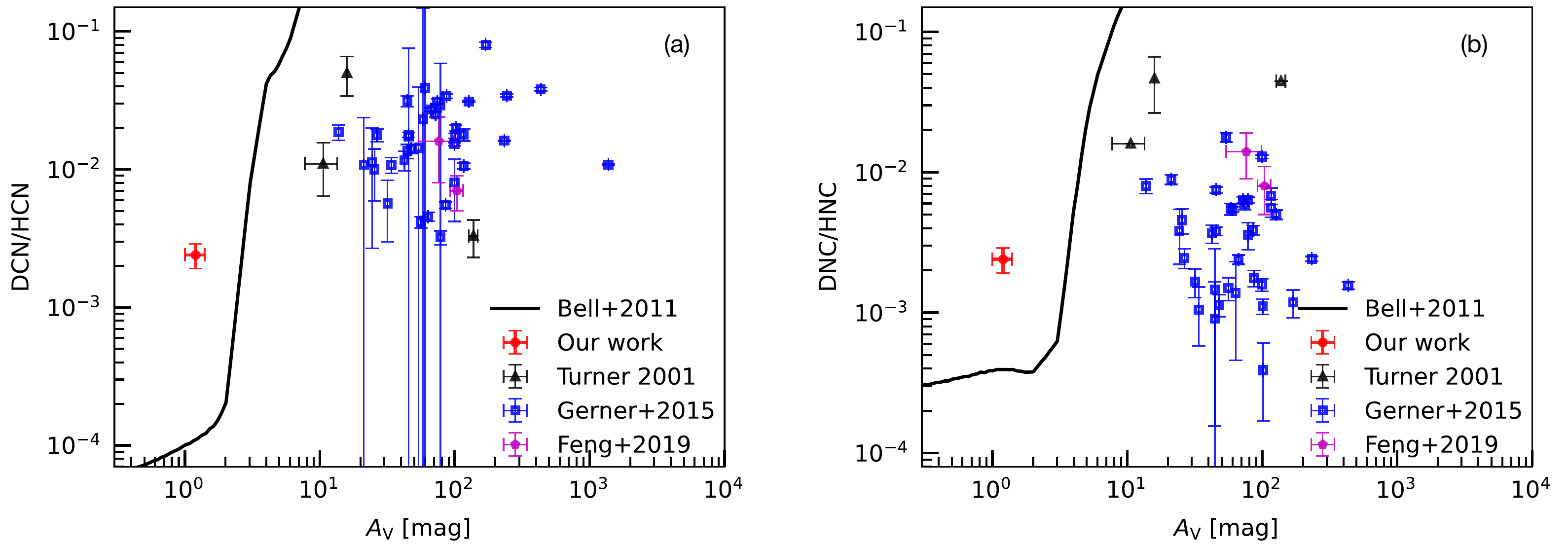}
\vspace{-1.5em}
\caption{(a): The comparison of the observed abundance ratios of DCN/HCN at different $A_{\rm V}$ in the literature, the values of $A_{\rm V}$ in dense cores are converted through $A_{\rm V} = N_{\rm H}/(1.89\times10^{21})$~mag. The red dot represents our measurement. Black triangles represent data taken from dark clouds \citep{Turner2001}, blue squares denote values from massive star-forming regions \citep{Gerner2015}, and purple points are taken from infrared dark clouds \citep{Feng2019}. The black solid lines represent the model predictions by \citet{Bell2011} without turbulent diffusion. The cloud density profile is defined as $n_{\rm H}$ = $n_0$r$_0$/r (where $n_0$ = 888~cm$^{-3}$, r$_0$ = 1.11~pc is the cloud size) in their model. (b) The same as (a) but for DNC/HNC.
}
\label{fig:fig2}
\end{figure*}

Figure \ref{fig:fig2} shows the observed DCN/HCN and DNC/HNC ratios at different $A_{\rm V}$ from our work and literature, overlaid by model predictions from \citet{Bell2011}. Compared with the values in the literature, our observation has the lowest volume density and $A_{\rm V}$ (see Appendix \ref{sec:density} and \ref{sec:column density}). 
The DCN/HCN abundance ratio in our work is almost an order of magnitude lower than the average values in massive star-forming regions \citep{Gerner2015} and cold, dense cores \citep{Turner2001,Feng2019,Mo2025}. 
The DNC/HNC abundance ratio in our work is much lower than cold, dense cores, but does not show much difference from those measurements in massive star-forming regions. 

The observed deuterium fractions are an order of magnitude higher than the model prediction, although the gas density adopted in their model exceeds that of our target by more than a factor of two. This substantial discrepancy cannot be accounted for by turbulent diffusion, which mixes molecules across different depths within molecular clouds \citep{Xie1995}, given the absence of any evidence for a dense, high-extinction region along the line of sight or within our field of view. 
We note that the model assumed a much lower gas temperature ($T_{\rm k}=10$~K); a higher temperature would further reduce the deuterium fraction \citep{Roueff2007}. 

\section{Discussion on the high deuterium fraction}\label{sec:chemistry}

Previous absorption observations toward PKS 1830-211 at {\it z}=0.89 reported high deuteration \citep[ND, NH$_2$D, and HDO,][]{Muller2020} as well as many high-{\it J} transitions of dense gas tracers \citep{Muller2011,Muller2013}. However, this absorber is known to be inhomogeneous and time-variable, containing denser sub-components in addition to the lower-density gas \citep{Henkel2009}. The absorption of dense gas tracers can originate from denser substructures along the line of sight even if the averaged density is in the order of a few $10^3$~cm$^{-3}$ \citep{Muller2020}. In contrast, our detection lies at much lower $n_{\rm H_2}$ and $N_{\rm H_2}$ regime and is difficult to explain under current astrochemical models.

The derived physical parameters of the translucent cloud suggest that the observed cloud is gravitationally unbound unless confined by external pressure and/or a very strong magnetic field (Appendix \ref{sec:dynamics}). We propose that the elevated deuterium fraction detected in the low-density cloud may reflect its evolutionary state: the translucent cloud is a transient, dynamically evolving structure that originates from the dissipation of unbound dense molecular clouds, a concept discussed more than two decades ago \citep{Price2003}. Magnetohydrodynamic (MHD) simulations have shown that dense cores ($n_{\rm H} \geq 10^4$ cm$^{-3}$) do not always collapse to form stars; instead, a substantial fraction ($\geq$20\%) of gravitationally unbound dense cores that fail to form stars could disperse into the low-density medium over a few $10^5$ yr \citep{Nakamura2005,Smullen2020,Offner2022}. In this scenario, molecular abundances in diffuse and translucent clouds may be enhanced relative to those characteristic of molecular clouds that form directly from diffuse atomic gas and evolve toward a quasi-equilibrium state \citep{Price2003}. 

\begin{figure}
\centering
\includegraphics[width=0.49\textwidth]{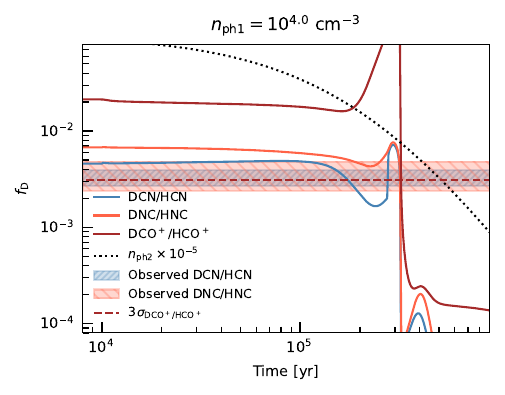}
\vspace{-1.5em}
\caption{Predicted DCN/HCN (blue), DNC/HNC (orange), and DCO$^+$/HCO$^+$ (red) ratios as a function of time from dispersal cloud models. The density on the top represents the maximum density in the collapsing phase. The black dotted line denotes the density function in the dispersal phase. The blue and orange shadowed regions are the observed DCN/HCN and DNC/HNC ratios with 1$\sigma$ uncertainties, and red dashed lines represent the 3$\sigma$ upper limits of DCO$^+$/HCO$^+$ from observations.  
}
\label{fig:fig3}
\end{figure}

To test the above hypothesis, we constructed a ``two-phase'' toy model (see Appendix \ref{sec:models}) to simulate the evolution of a molecular cloud, beginning with free-fall collapse and followed by a dispersal phase. Figure \ref{fig:fig3} shows the evolution of the deuterium fractions (DCN/HCN, DNC/HNC, and DCO$^+$/HCO$^+$) during the dispersal phase. During this transient phase, the deuterium fraction can remain above the observational detection limits before declining rapidly once the gas density falls below a few $\sim10^2$~cm$^{-3}$, primarily because of photodissociation. Therefore, if the processes described above enrich deuterated molecules during dissipation, the deuterium fraction in translucent clouds is expected to remain elevated, exceeding those predicted by a quasi-equilibrium state. 

\section{Conclusion}\label{sec:conclusion}

For the first time, we have detected deuterated molecules (DCN and DNC) in a Galactic translucent cloud ($A_{\rm V}=1.2\pm0.2$~mag, $n_{\rm H_2}=398\pm22$~cm$^{-3}$). The unexpected high deuterium fraction represents a significant departure from established chemical models. Although the mechanism by which deuterated molecules accumulate to such abundances remains unclear, these results provide a hint that the chemistry in translucent clouds may be different from the canonical picture that diffuse and translucent cloud evolve from diffuse atomic gas. Given that the cloud is not gravitationally bounded, we consider the scenario that the translucent cloud is a transient phase that originate from a dispersed dense core, which may help explain the overabundance of deuterated molecules. Future studies based on larger samples obtained with NOEMA and ALMA, together with more detailed chemical simulations, should provide further insight into deuterium fractionation and cloud evolution.

\begin{acknowledgements}
We thank the staff and operators at IRAM for advice and assistance with observations and data reduction.
TGB acknowledges support from the Leading Innovation and Entrepreneurship Team of Zhejiang Province of China (Grant No. 2023R01008). 
DL acknowledges support from the New Cornerstone foundation. 
S.F. acknowledges support from the National Key R\&D program of China grant (2025YFE0108200) and National Science Foundation of China (12373023). 
This work is based on observations carried out under project numbers W20BB and E25AE with the IRAM NOEMA Interferometer. IRAM is supported by INSU/CNRS (France), MPG (Germany) and IGN (Spain). 
\end{acknowledgements}

%
%

\bibliographystyle{aa}
\bibliography{reference} 



\begin{appendix} 
\nolinenumbers

\section{Molecular transitions and fitting results}\label{sec:table}

Table \ref{tab:tab1} lists the frequencies, optical depths, and column densities at $-0.9$ km\,s$^{-1}$ component for observed molecular transitions. The column densities are derived through equation (32) in \citet{Mangum2015} and the excitation temperature is adopted as 2.73~K, given the gas density (Appendix \ref{sec:density}) is far below the critical density of the concerned transitions. We note that the hyperfine transitions of HNC and DNC cannot be distinguished under the current spectral resolution; the column densities were derived by integrating the optical depths. The derived HNC column density is consistent with \citet{Lucas1998}.

\begin{table}[!ht]
\caption{Rest frequencies, optical depths, and column densities at $-0.9$ km\,s$^{-1}$ component for observed molecular transitions.}
\label{tab:tab1}   
\centering          
\begin{tabular}{c c c c}     
\hline\hline  
 Molecules& Frequency & $\tau$ & $N_\mathrm{col}$  \\
  & GHz & & 10$^{12}$ cm$^{-2}$\\
\hline
H$^{13}$CO$^+$ & 86.7542884 & 0.104$\pm$0.001 & 0.123$\pm$0.003 \\
DCO$^+$ & 72.0393042 & $<$0.008 (3$\sigma$) & $<$0.028  \\
H$^{12}$CN & 88.6318475 & 5.7$\pm$0.2\tablefootmark{a} & 18.1$\pm$0.6  \\
DCN & 72.414933 & 0.011$\pm$0.001\tablefootmark{a} & 0.059$\pm$0.011 \\
HN$^{12}$C & 90.663568 & 2.11$\pm$0.03 & 4.15$\pm$0.06   \\
DNC & 76.305727 & 0.005$\pm$0.001 & 0.015$\pm$0.005 \\
\hline                  
\end{tabular}
\tablefoot{\tablefoottext{a}{Optical depth of {\it J}=1-0, {\it F}=2-1 transition.}}
\end{table}

\section{The gas volume density at $-0.9$ km\,s$^{-1}$ component}\label{sec:density}

The physical parameters (e.g., density, temperature) are the key to understanding the chemical processes of the molecular gas. We use the non-LTE radiative transfer code {\sc radex} \citep{Van2007} to constrain the gas volume density, with the archival observational data from $^{13}$CO ($J$=1--0 and 2--1) and NH$_3$ (1,1) and (2,2) \citep{Liszt1998,Liszt2006}. 
The input free parameters in {\sc radex} modeling are gas kinetic temperature ($T_{\rm k}$), H$_2$ volume density ($n_{\rm H_2}$), $^{13}$CO column density ($N_{\rm ^{13}CO}$), and NH$_3$ column density ($N_{\rm NH_3}$). The linewidth is fixed to 1~km~s$^{-1}$. The collision rate and coefficients of $^{13}$CO and NH$_3$ with H$_2$ are adopted from \cite{Yang2010} and \cite{Loreau2023}. We use the Markov Chain Monte Carlo (MCMC) method within the $emcee$ code \citep{Foreman2013} to sample the posterior probability distributions of the above free parameters, in which the likelihood function of the posterior probability function is defined as:
\begin{equation}
\rm ln \ p = -\frac{1}{2} \sum_i \left [ \frac{\left ( \tau_{obs}^i - \tau_{model}^i \right )^2}{{\sigma^i_{obs}}^2} + ln \left ( 2\pi {\sigma^i_{obs}}^2 \right ) \right ],
\end{equation}
where the ${\rm \tau_{obs}^i}$ and ${\rm \sigma^i_{obs}}$ are the observed optical depth and its uncertainty of the $i$th transition, respectively. ${\rm \tau_{model}^i}$ is the modeled optical depth by {\sc radex}. 

\begin{figure}
\centering
\includegraphics[width=0.5\textwidth]{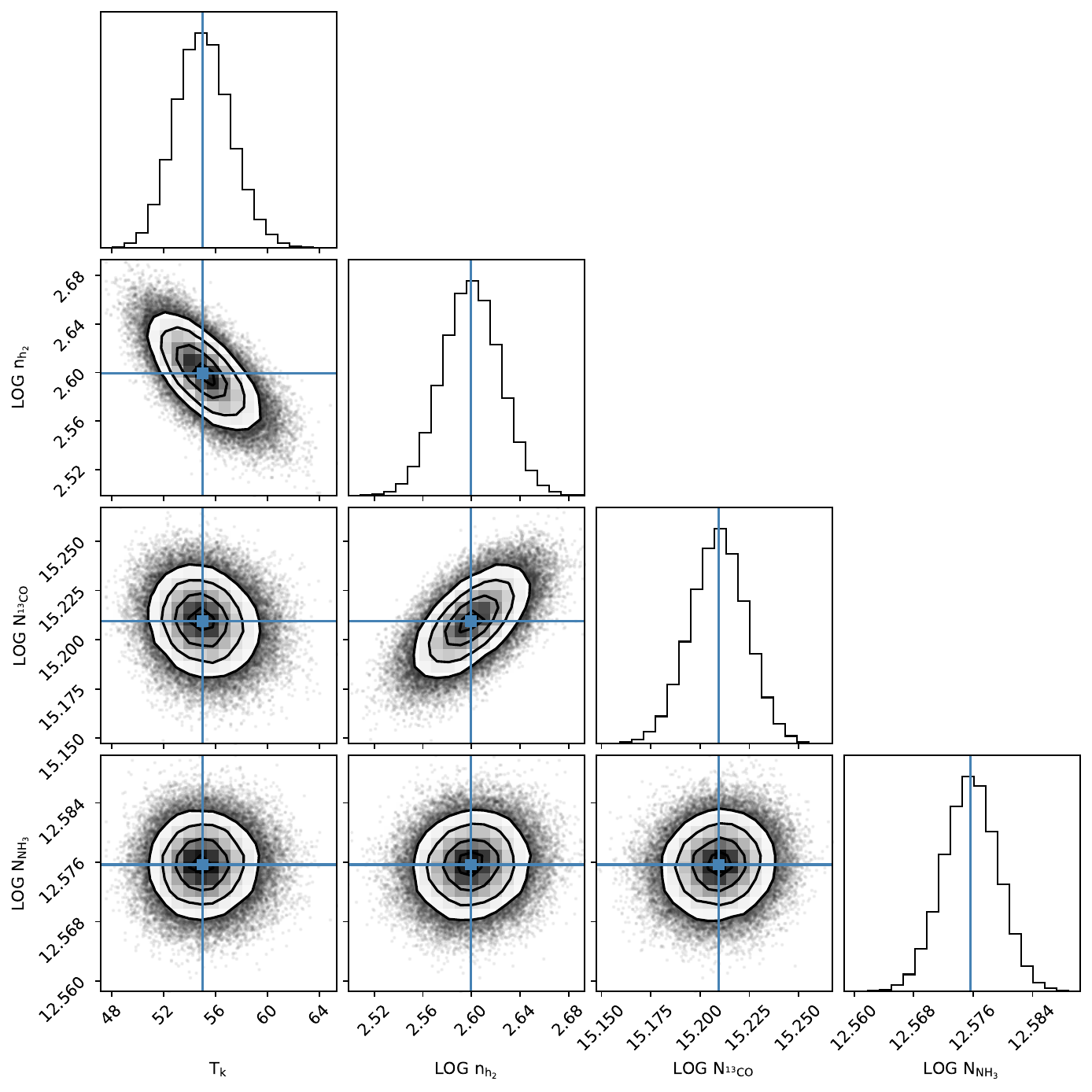}
\caption{
The probability distribution of free parameters $T_{\rm k}$, $n_{\rm H_2}$, $N_{\rm ^{13}CO}$, and $N_{\rm NH_3}$.
}
\label{fig:cornermap}
\end{figure}

The MCMC fitting shows good convergence in the sampled parameter space (Fig.~\ref{fig:cornermap}). The resultant $T_{\rm k}$, $n_{\rm H_2}$, $N_{\rm ^{13}CO}$, and $N_{\rm NH_3}$ are 55$\pm$2\,K, 398$\pm$22\,cm$^{-3}$, $(1.6\pm0.1)\times10^{15}$\,cm$^{-2}$, and $(3.76\pm0.03)\times10^{12}$\,cm$^{-2}$, respectively. The results of column densities are consistent with previous estimates by \citet{Liszt1998} and \citet{Liszt2006}. While the pressure ($n_{\rm H_2}T_{\rm k}$) is more than twice higher than previous estimates toward the same sightline \citep{Lucas1996}, the latest collision rate coefficient makes the new calculation more robust.

\section{The gas column density at $-0.9$ km\,s$^{-1}$ component}\label{sec:column density}

The stellar extinction statistics give a value of $A_{\rm V}$=0.85$\pm$0.15 mag for the $-0.9$ km\,s$^{-1}$ component and 2.2$\pm$1.0 mag for the $-2.5$ km\,s$^{-1}$ component \citep{Lucas1998}. This is reasonably consistent with the total column density derived from the {\it Planck-Herschel} dust map \citep[$A_{\rm V} \sim 2.9$ mag,][]{Lada2017}. However, since the measurements of stellar extinction and dust thermal emission do not exactly fit the spatial scale of NOEMA observations, the actual column density along the sightline may deviate from the above estimation if spatial fluctuation is significant. Here we estimate the total H column density through two independent ways: CH observations at a similar angular resolution as NOEMA and $^{13}$CO measurements from Appendix \ref{sec:density}. 

The hyperfine structure line of CH within the ground state $\Lambda$-doublet $^2\Pi_{1/2}$, $J$ = 1/2 energy level at 3335~MHz was observed with the Karl G. Jansky Very Large Array (VLA) in B-configuration in Oct. 2021 (Project ID: 21B-120, PI: Gan Luo). The integration time is $\sim$15 mins, and the spectral resolution of VLA is 3.91 kHz. The raw data were calibrated using the standard pipeline with the Common Astronomy Software Applications \citep[CASA, version 5.6.1,][]{CASA2007}. The imaging of the calibrated visibilities was performed using the {\it tclean} algorithm with Briggs weighting (robust = 0.5). The beam size of the final cleaned images is $2.9''\times1.8''$ (position angle PA = 73$^\circ$) at 3335 MHz, which is comparable to the NOEMA observations.

\begin{figure}
\centering
\includegraphics[width=0.5\textwidth]{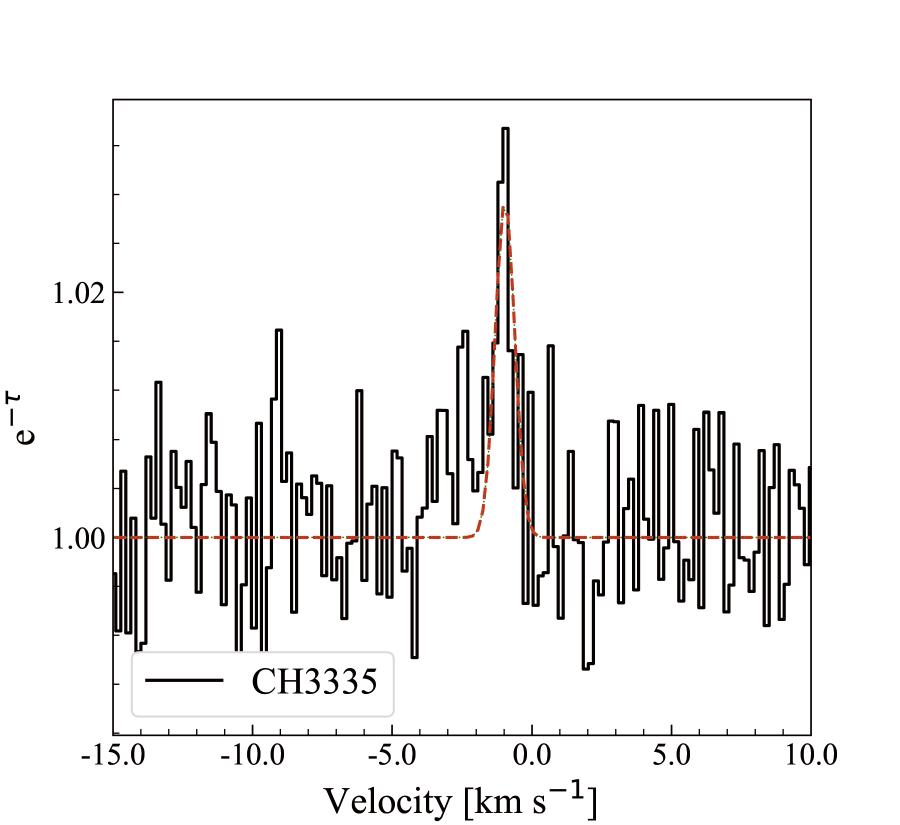}
\caption{
The normalized spectra (black solid line) and fitting result (red dashed line) of CH 3335 MHz toward J0418+3801.
}
\label{fig:spec_ch}
\end{figure}

Figure \ref{fig:spec_ch} shows the normalized spectra of the CH 3335~MHz, in which only the $-0.9$~km~s$^{-1}$ component was detected. The emission profile against a bright background ($T_{\rm bg} >10^4$~K) directly reflects the nature of level inversion (negative $T_{\rm ex}$) of CH 3335~MHz toward J0418+3801, as it is observed in many other environments \citep{Rydbeck1976,Jacob2021,Tang2021,Luo2025}. The optical depth derived from the emission profile is $\tau = -0.027\pm0.004$, which is in good agreement with previous NRAO 43m single-dish observations \citep[$\tau\sim0.03$,][]{Liszt2002}. Assuming that spatial fluctuations can be neglected within the single-dish beam area, we derive the excitation temperature of CH by combining previous single-dish and VLA observations, obtaining a value of $T_{\rm ex} = -2.4 \pm 1.2$ K, which is consistent with recent measurements in translucent clouds \citep{Tang2021}. The CH column density is therefore $(1.9\pm1.0)\times10^{13}$ cm$^{-2}$. Since the abundance of CH is fairly constant in a wide H$_2$ column density range \citep[$N_{\rm CH}/N_{\rm H_2}=3.5\times10^{-8}$,][]{Sheffer2008}, the estimated H$_2$ column density is $(5.6\pm2.9)\times10^{20}$ cm$^{-2}$. 

On the other hand, assuming $^{12}$C/$^{13}$C=74\footnote{The value could be overestimated due to fractionation \citep{Liszt2007,Colzi2020,Sipila2023}.} and $^{12}$CO/H$_2$ = 1.0$\times$10$^{-4}$, we can obtain $N_{\rm H_2}$ from the derived $^{13}$CO column density, resulting a higher $N_{\rm H_2} = (1.2\pm0.1)\times10^{21}$ cm$^{-2}$. 
Adopting H\,{\sc i} column density in the range of $(4.57\pm0.32)\times10^{20}$ cm$^{-2}$ from the 21-SPONGE H\,{\sc i} Absorption Line Survey \citep{Murray2018}, the total H column densities at $-0.9$ km\,s$^{-1}$ component are $1.6\times10^{21}$ and $2.8\times10^{21}$ cm$^{-2}$ from the two estimates, corresponding to $A_{\rm V}$ values of 0.85 and 1.5~mag (assuming the canonical conversion factor $N_{\rm H}$/$A_{\rm V}$ = $1.89\times10^{21}$ atoms\,cm$^{-2}$\,mag$^{-1}$). In this paper, we use an averaged $A_{\rm V} =1.2\pm0.2$~mag for the $-0.9$~km~s$^{-1}$ component. 

\section{Dynamical state of the translucent cloud}\label{sec:dynamics}

The derived parameters in Appendix \ref{sec:density} and \ref{sec:column density} suggest a cloud depth along the line-of-sight is $L\approx0.8~{\rm pc}$. Assuming a uniform spherical cloud with ($n_{\rm H_2}=400~{\rm cm^{-3}}$), ($T_k=55~{\rm K}$), and diameter ($L=0.8~{\rm pc}$), we obtain ($M\approx7.4~M_\odot$). The observed H$^{13}$CO$^+$ linewidth of (1~km\,s$^{-1}$) implies a non-thermal velocity dispersion of ($\sigma_{\rm nt}\approx$0.4~km\,s$^{-1}$). The corresponding thermal energy is:
\begin{align}
E_{\rm th}
&=
\frac{3}{2}\frac{M}{\mu m_{\rm H}}k_{\rm B}T_k
=
4.3\times10^{43}~{\rm erg},
\end{align}
and the turbulent energy is:
\begin{align}
E_{\rm turb}
&=
\frac{3}{2}M\sigma_{\rm nt}^2
=
3.7\times10^{43}~{\rm erg}.
\end{align}
The gravitational energy ($E_{\rm grav}=-\frac{3}{5}\frac{GM^2}{R}=-7.1\times10^{42}$) erg is an order of magnitude lower than the sum of thermal and turbulent energies, suggesting the cloud is gravitationally unbounded. To keep the cloud pressure confined, one needs a B-field strength of ($B>13~\mu{\rm G}$) ($P_B=\frac{B^2}{8\pi}=2/3\frac{E_{\rm th}+E_{\rm turb}}{V}$) to balance the gas pressure (assuming the magnetic field provides external pressure). However, the measurements of magnetic field through the Zeeman effect are much lower in such a density and column density range, usually a few $\mu{\rm G}$ \citep{Crutcher2012}. Note that even considering a different morphology, for example, an infinite slab, it would still require a B-field ($B>11~\mu{\rm G}$) to balance the pressure.

\section{Our toy models}\label{sec:models}

To model the observed deuterium fraction, we perform ``two-phase'' model tests with {\sc uclchem} \citep{Holdship2017}, which consist of an initial isothermal collapse ($T = 10$ K) phase (``phase1'') and are followed by a dispersal phase (``phase2''). The ``phase1'' evolves from a diffuse cloud with density $n_{\rm ini} = 10^2$ cm$^{-3}$, reaching to $n_{\rm ph1}$ = 10$^4$~cm$^{-3}$ under free-fall collapse \citep[see][for a detailed description of the density profiles in collapse models]{Priestley2018}. Then, it is followed by the ``phase2'' stage, in which the gas density $n_{\rm ph2}$ decreases as a function of time following the equation from \citet{Price2003}:
\begin{equation}
    n_{\rm ph2} = n_{\rm ph1}/(1+\alpha t/t_0)^3,
\end{equation}
where $t_0/\alpha$ is the parameter that can be adjusted to determine the time to reach the density of diffuse clouds. We take it to be $5\times10^4$ yr so that the cloud would disperse to a diffuse cloud at a few $10^5$ yr.

The deuterium chemical network of this work adopts the same networks as described in \citet{Majumdar2017}. To reduce computational cost, we consider only species with molecular weight $<$ 66 (i.e., carbon-chain and complex organic molecules with more than 5 carbon atoms are excluded). Since {\sc uclchem} cannot include both ortho- and para-H$_2$ in the network, we only consider para-H$_2$ reactions as para-H$_2$ is the dominant one in cold gas\footnote{We note here that the deuterium network employed in our model is incomplete; we only use it for the demonstrative purpose of a dispersal core scenario.} \citep{Flower1984}. The chemical network contains a total of $\sim800$ species and $\sim 3\times10^4$ reactions. We follow the same recipe as \citet{Luo2024a} when running the chemical models. The external FUV field is set to $\chi/\chi_0 = 1$, normalized to the spectral shape given in \citep{Draine1978}. The cosmic-ray ionization rate per H$_2$ ($\zeta_2$) is adopted to be $1.7\times10^{-17}$ s$^{-1}$. The size of the cloud is 0.8 pc all the time. 

Figure \ref{fig:fige1} shows the modeled abundance ratios of DCN/HCN, DNC/HNC, and DCO$^+$/HCO$^+$ as a function of time during the dispersal phase at different $n_{\rm ph1}$ hypotheses. At $n_{\rm ph1}=10^3$~cm$^{-3}$, the abundance ratios of DCN/HCN and DNC/HNC are more than an order of magnitude lower than the observed ones all the time. 
Although the models can predict a high deuterium fraction in the dispersal phase, we note that the DCO$^+$/HCO$^+$ abundance ratio is too high and may be overestimated before the dispersal of the core, as we do not observe such a high value in warm environments. This might be related to the incomplete network employed in our models, as the DCO$^+$/HCO$^+$ abundance ratio is sensitive to the ortho- to para-H$_2$ ratio \citep{Shingledecker2016}. We emphasize that the parameters we used in our toy models are not the only recipe that can yield this solution. The density functions in two phases, the time-scale for which the cloud starts to disperse, the deuterium network in our model, assumed physical parameters, etc, all these will change the behavior of the model predictions. We cannot take all the things into account with simple toy models. Nevertheless, the scope of the toy model is not to find a prediction that matches perfectly with all observables, but to demonstrate that the scenario of a dispersal cloud can lead to the observed deuterium fraction. More detailed chemical simulations in the future will be helpful to understand the deuterium fractionation in the translucent cloud.

\begin{strip}
\centering
\includegraphics[width=1.0\textwidth]{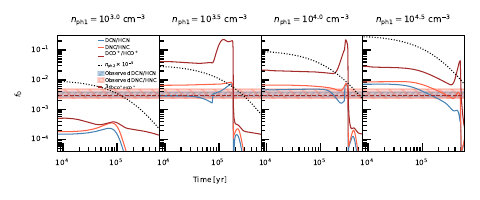}
\captionof{figure}{The same as Fig.~\ref{fig:fig3} but for different $n_{\rm ph1}$ as labeled at the top of each panel. 
}
\label{fig:fige1}
\end{strip}
\par\noindent\mbox{}\par

\end{appendix}
\end{document}